\documentclass[onecolumn,showpacs,showkeys,superscriptaddress]{revtex4-2}

\pdfoutput=1
\usepackage{booktabs}
\usepackage{multirow}
\usepackage{natbib}
\usepackage{physics}
\usepackage{dcolumn}
\usepackage{bm}
\usepackage{amsmath}
\usepackage{mathtools}
\usepackage{amsfonts}
\usepackage{pbox}
\usepackage{color}
\usepackage[dvipsnames]{xcolor}
\usepackage{tabularx}
\usepackage{subfigure}
\usepackage{graphicx}
\usepackage{psfrag}
\usepackage{lipsum}
\usepackage{enumitem}
\usepackage{changes}
\usepackage{array}
\usepackage{CJK}
\usepackage[english]{babel}

\usepackage[colorlinks = true,
            linkcolor = blue,
            urlcolor  = blue,
            citecolor = blue,
            anchorcolor = blue]{hyperref}
\usepackage{blindtext}
\usepackage[utf8]{inputenc}
\usepackage{graphicx}
\usepackage{placeins}
\usepackage{siunitx}
\usepackage{amssymb}
\usepackage{mathtools}
\usepackage{physics}
\usepackage{tikz}
\usepackage[titletoc]{appendix}
\usepackage{listings}
\usepackage{float}
\usepackage{braket}

\usepackage[capitalize]{cleveref}
\usepackage{soul}
\usepackage{array}

\begin{document}

\title{Effects of anisotropy in an anisotropic extension of $w$CDM model}
\author{Vikrant Yadav}
\email{vikuyd@gmail.com}
\affiliation{School of Basic and Applied Sciences, Raffles University, Neemrana - 301705, Rajasthan, India}

\author{Santosh Kumar Yadav}
\email{sky91bbaulko@gmail.com}
\affiliation{School of CS \& AI, SR University, Warangal-506371, Telangana, India}

\author{Rajpal}
\email{rajpal05041985@gmail.com}
\affiliation{School of Basic and Applied Sciences, Raffles University, Neemrana - 301705, Rajasthan, India}

\begin{abstract}
In this paper, we derive observational constraints on an anisotropic $w$CDM model from datasets including Baryonic Acoustic Oscillations (BAOs), Cosmic Chronometer (CC), Big Bang Nucleosynthesis (BBN), Pantheon Plus (PP) compilation of Type Ia supernovae,  and SH0ES Cepheid host distance anchors. 
We find that anisotropy is of the order $10^{-13}$, and its presence in the $w$CDM model reduces the $H_0$ tension by $\sim 2\sigma$ and $\sim 1\sigma$ in the analyses with BAO+CC+BBN+PP and BAO+CC+BBN+PPSH0ES data combinations, respectively. In both analyses,  the quintessence form of dark energy is favored at 95\% CL.

\end{abstract}


\maketitle
\section{Introduction}



\label{sec:intro}
Our Universe is expanding with an accelerated rate of expansion as observed from type Ia supernovae (SNe Ia) observations~\cite{SupernovaSearchTeam:1998fmf,SupernovaCosmologyProject:1998vns}. Later on, various other measurements such as the large-scale structure (LSS), the Baryonic Acoustic Oscillations (BAOs), and the cosmic microwave background (CMB) supported this observation. Several theoretical frameworks have been proposed/investigated in the literature to help us comprehend the dynamics of the Universe. But, among all these, the $\Lambda$CDM model has proven to be the simplest mathematical framework widely accepted by the research community and referred to as the standard model of cosmology. It is composed of two major components: cold (non-relativistic) dark matter (CDM) which is the reason behind the structure formation, and dark energy (DE) in the form of cosmological constant ($\Lambda$) which causes the late time accelerated expansion of Universe. This $\Lambda$CDM paradigm provides an excellent fit across a broad range of scales and epochs~\cite{Planck:2018vyg,DES:2018paw,Blomqvist:2019rah,DES:2021wwk, Riechers:2022tcj}, and successfully describes late-time accelerated expansion of the Universe~\cite{Copeland:2006wr,Bamba:2012cp}.
Despite the excellent fit to the current available cosmological datasets, the model faces several theoretical and observational challenges. For instance, the nature of DE is not known accurately so far. Within the standard $\Lambda$CDM scenario, DE is regarded as the cosmological constant in its most basic form, lacking any solid physical foundation. The exact nature of dark matter remains unknown except for its typical gravitational interactions with other components. Also, there is no concrete explanation of the `coincidence problem' which states: why, despite having very distinct cosmic evolutions, do the dark matter and the DE densities have the same order at present?
 Is this coincidence indicating the possibility of interaction between the dark sector components? Up to what extent the cosmological principle has been tested? Is the Universe homogeneous and isotropic at cosmic scales? 

Further, a few indications in the observations point towards the need to expand the $\Lambda$CDM model to account for the growing conflicts between measurements made at early (high redshifts) and late (low redshifts) Universe~\cite{Verde:2019ivm,Clark:2021hlo}. The Hubble constant, $H_0$, which represents the Universe's current rate of expansion, has the greatest statistically significant tension. The Hubble tension appears when we compare the value of $H_0$ predicted by CMB measurements within the $\Lambda$CDM framework and the direct local distance ladder measurements, that is, the one estimated by the Cepheid calibrated SNe Ia. In particular, the Hubble tension is referred to as the disagreement at $5\sigma$ between the latest SH0ES  (Supernovae and $H_0$ for the equation of state (EoS) of DE) collaboration \cite{Riess:2021jrx} constraint, $H_0^{\rm R22}=(73.04 \pm 1.04){\rm \,km\,s^{-1}\,Mpc^{-1}}$ at 68\% confidence level (CL), based on the supernovae calibrated by Cepheids, and the {\it Planck} collaboration~\cite{Planck:2018vyg} value, $H_0=\left(67.27\pm0.60\right){\rm \,km\,s^{-1}\,Mpc^{-1}}$ at 68\%  CL. 

This disparity could indicate the existence of novel physics outside of the $\Lambda$CDM cosmology \cite{Freedman:2017yms,Mortsell:2018mfj,Knox:2019rjx,DiValentino:2021izs,Vagnozzi:2021gjh,Hu:2023jqc}. In the literature, many extensions of $\Lambda$CDM have been suggested to resolve the $H_0$ tension. These extensions include but not limited to: DM-DE interactions \cite{Pan:2019gop,Kumar:2016zpg,Kumar:2017dnp,Yang:2018uae,Kumar:2019wfs,Kumar:2021eev,Kaeonikhom:2022ahf};
decaying DM \cite{Bringmann:2018jpr,Kumar:2018yhh,Yadav:2019jio,Yadav:2023qfj}; 
introducing Early DE \cite{Poulin:2018cxd,Niedermann:2020dwg,Hill:2020osr,Poulin:2023lkg,Reeves:2022aoi}; 
and introducing a sign-switching DE at intermediate redshifts ($z\sim 2$)~ \cite{Akarsu:2019hmw,Akarsu:2021fol,Akarsu:2022typ,Akarsu:2023mfb,akarsu2024lambdarm}. The current status on $H_0$ tension and possible solutions can be found in recent review articles~\cite{Perivolaropoulos:2021jda,Abdalla:2022yfr,Vagnozzi:2023nrq,Freedman:2023jcz}

The Cosmological Principle (CP), which asserts that the Universe is statistically homogenous and isotropic in space and matter on vast scales $(\sim 100\rm Mpc)$, is one of the fundamental tenets of $\Lambda$CDM cosmology. Mathematically, such a Universe is described by the Friedmann-Lemaître-Robertson-Walker (FLRW) space-time metric, in which all three of the metric's spatial orthogonal components are functions of cosmic time ($t$) exclusively. This is the primary space-time metric that facilitates the creation of the effective and remarkably predictive standard model of cosmology. However, observational data suggest that there are slight fluctuations in CMB intensities originating from various directions of sky ~\cite{Yeung:2022smn}. Gamma Ray Bursts (GRBs), Quasars, Galaxies, and SNe Ia are among the recent findings that suggest the Universe may be anisotropic and provide substantial observational evidence (see \cite{Hu:2020mzd} and references therein). There are other interesting pieces of evidence, that have questioned the validity of CP, for instance, a piece of strong evidence for a violation of the CP of isotropy is found by the authors in \cite{Fosalba:2020gls} after analyzing the Planck Legacy temperature anisotropy data. This violation is likely to represent a statistical fluctuation of order $\sim 10^{-9}$. Also, see the recent and interesting review \cite{Aluri:2022hzs} where the authors emphasize differences and synergies that truly stimulate additional research in this field as they detail existing observational clues for departures from the predictions of CP. The spatially homogeneous and anisotropic Universe is expressed mathematically by homogeneous and anisotropic metrics and corresponding models are commonly known as Bianchi-type models \cite{Watanabe:2009ct,Kanno:2010nr}. The simplest one among these is the Bianchi type I.

The empirical limitations on the Bianchi type I model might offer a platform for evaluating the precision of FLRW frameworks characterizing the current era of the Universe. Various anisotropic generalizations of the standard $\Lambda$CDM framework have been investigated in recent years. In \cite{Akarsu:2019pwn}, the authors analyze an anisotropic scenario with the CMB and BAO data by fixing the drag redshift as $z_{\rm d}=1059.6$ and the last scattering redshift as $z_* = 1089.9$ and constrain the anisotropy parameter, $\Omega_{\rm \sigma0}\lesssim10^{-15}$. A similar investigation is done in \cite{Akarsu:2021max} by considering anisotropic expansion and curvature together from different data sets and discovering little present-day expansion anisotropy. Other anisotropic models are also investigated by using different data sets in different contexts, see \cite{Amirhashchi:2018bic,Amirhashchi:2018nxl,Amirhashchi:2020qep,yadav2021transitioning, yadav2021constraining, Rahman:2021mti, Bhardwaj:2023wxt}. In \cite{Constantin:2022dtj}, a theoretical approach to a spatially homogeneous Universe with late-time anisotropy is discussed. Most recently, in \cite{Yadav:2023yyb} an anisotropic extension, $\Lambda$CDM + $\Omega_{\rm \sigma0}$ is investigated to derive CMB independent constraints on the Hubble constant with BAO, BBN, CC, Pantheon Plus and SH0ES data. Following \cite{Yadav:2023yyb}, we are motivated here to derive constraints on two simplest extensions of vanilla $\Lambda$CDM paradigm: (i) an isotropic extension by taking a constant EoS of DE instead of a cosmological constant, namely the $w$CDM model; (ii) an anisotropic extension, namely $w$CDM + $\Omega_{\rm \sigma0}$ by adding expansion anisotropy to the $w$CDM model.
It is well known that the nature of DE could be investigated through its EoS parameter. As stated above, the fundamental nature of DE is unknown and there is no strong physical basis for assuming DE in the form of a cosmological constant whose density remains constant even in an expanding background. In this paper, we have assumed a constant EoS parameter of DE. Our primary goal in this work is to analyze the possible effects of anisotropy on the constant EoS parameter of DE and the Hubble constant $H_0$. For this, we derive the constraints on the free parameters of the considered models using recent data and compare our results with the results obtained in \cite{Yadav:2023yyb} for the $\Lambda$CDM and $\Lambda$CDM + $\Omega_{\rm \sigma0}$, respectively.  

The structure of the paper is as follows: In Section \ref{sec2}, we describe the mathematical details of the considered scenarios. In Section \ref{sec3}, we provide a brief overview of the data sets and methodology used for the analysis of the models. In Section \ref{sec4}, we present the constraints on the parameters under study and discuss the results. In Section \ref{sec5}, we conclude with the main findings of our study.

\section{GOVERNING EQUATIONS AND MODELS}\label{sec2}
We investigate an extension of FLRW spacetime, the Bianchi type I metric, with three orthogonal directions of different scale factors, given by
\begin{equation}\label{metric}
    \text{d}s^2=-\text{d}t^2 + A^2\text{d}x^2 + B^2\text{d}y^2 + C^2\text{d}z^2,
\end{equation} 
where $A,B,$ and  $C$ are functions of cosmic time $t$ only and represent the scale factors along the principal axes $x, y$ and $z$, respectively. Further, we define the average expansion scale factor: $a=(ABC)^{\frac{1}{3}}$, and the average Hubble parameter:  $H=\frac{\dot{a}}{a}=\frac{\left(H_x + H_y+H_z\right)}{3}$, where the corresponding directional Hubble parameters along the principal axes $x,y$, and $z$ are being defined as $H_x =\frac{\dot{A}}{A}$, $H_y =\frac{\dot{B}}{B}$, and $H_z =\frac{\dot{C}}{C}$ respectively.

The Einstein field equations in GR read as
\begin{equation}\label{fieldeqn}   
G_{\mu\nu}\equiv R_{\mu\nu}-\frac{1}{2}g_{\mu\nu} R= 8\pi GT_{\mu\nu},
\end{equation}    
where the left side of the above expression shows the Einstein tensor $G_{\mu\nu}$, the Ricci tensor $R_{\mu\nu}$, the Ricci scalar $R$, and the metric tensor $g_{\mu\nu}$. Additionally, the right side shows Newton's gravitational constant $G$ and the energy-momentum tensor $T_{\mu\nu}$. Further, for a perfect fluid with energy density $\rho$ and pressure $p$, $T_{\mu\nu}$ takes the form 
\begin{equation}\label{emt}
 T^{\nu}_{\,\,\mu} = \text{diag} [-\rho, p, p, p].    
\end{equation}
As a result of the twice-contracted Bianchi identity ($G^{\mu\nu}_{\;\;\;;\nu}=0$), the Einstein field equations ~\eqref{fieldeqn} satisfy the conservation equation
\begin{equation}\label{ece}
 T^{\mu\nu}_{\;\;\;;\nu}=0.    
\end{equation}
In the case of the perfect fluid matter distribution, it reduces to 
\begin{equation}\label{ce}
 \dot{\rho}+3H(\rho+p)=0,    
\end{equation}
where the dot denotes its cosmic time($t$) derivative. The evolution of the energy density $\rho_i$ of a perfect fluid $i$ with pressure $p_i$ and constant EoS $w_i=p_i/\rho_i$ is provided by the continuity equation \eqref{ce}, which is as follows: 
\begin{equation}\label{eos}
\rho_i=\rho_{i0}a^{-3(1+w_i)},    
\end{equation}
where $\rho_{i0}$ stands for the current value of $\rho_{i}$, that is, at $a=a_0=1$. Here, $a=a_0=1$ signifies the present-day value of the cosmic scale factor $a$. From this point on, every quantity with the subscript 0  indicates its value in the present-day Universe. We consider that the Universe is made up of the standard cosmic fluids, namely the radiation (photons and neutrinos) whose EoS is $w_{\rm r}= p_r/\rho_r =\frac{1}{3}$, pressureless fluid (baryonic and cold dark matter) whose EoS is $w_{\rm m}= p_m/\rho_m=0$, and DE fluid with a constant EoS $w_{\rm de}$, to be fixed by observations in the analysis.  Further, assuming only gravitational interaction between these energy components, the continuity equation \eqref{ce} is satisfied by each component separately, and because of \eqref{eos}, this gives

\begin{equation}\label{sources}
\rho\equiv\rho_\text{r}+\rho_\text{m}+\rho_{\rm de}= \rho_{\text{r}0}a^{-4}+\rho_{\text{m}0} a^{-3}+\rho_{de0}a^{-3(1+w_{\rm de0})}
\end{equation}

Further, for the Bianchi type I metric~\eqref{metric}, the Einstein field equations~\eqref{fieldeqn} result into to the following set of differential equations:
\begin{align}
    \frac{\dot{A}}{A}\frac{\dot{B}}{B}+ \frac{\dot{B}}{B}\frac{\dot{C}}{C}+ \frac{\dot{A}}{A}\frac{\dot{C}}{C}&=8\pi G \rho \label{BI.1}, \\ 
    -\frac{\Ddot{B}}{B}-\frac{\Ddot{C}}{C}- \frac{\dot{B}}{B}\frac{\dot{C}}{C}&=8\pi G p \label{BI.2}, \\ 
      -\frac{\Ddot{A}}{A}-\frac{\Ddot{C}}{C}- \frac{\dot{A}}{A}\frac{\dot{C}}{C}&=8\pi G p \label{BI.3} , \\
        -\frac{\Ddot{A}}{A}-\frac{\Ddot{B}}{B}- \frac{\dot{A}}{A}\frac{\dot{B}}{B}&=8\pi G p \label{BI.4} . 
\end{align}
The shear scalar may be expressed as follows in terms of the directional Hubble parameters:
\begin{equation}\label{BIshear} 
\sigma^2 = \frac{(H_x -H_y)^2 + (H_y - H_z)^2 + (H_z -H_x)^2}{6}.
\end{equation} 
The equations~\eqref{BI.1}-\eqref{BI.4} can be recast as follows: 
\begin{align} 
3H^2-\sigma^2=8\pi G\rho,\label{hubble0}\\
     \dot{H}_x-\dot{H}_y+3H(H_x-H_y)=0,\label{hubble1} \\
      \dot{H}_y-\dot{H}_z+3H(H_y-H_z)=0,\label{hubble2} \\  
       \dot{H}_z-\dot{H}_x+3H(H_z-H_x)=0\label{hubble3}.    
\end{align} 
We derive the following shear propagation equation using these equations \eqref{hubble1}-\eqref{hubble3} and the time derivative of $\sigma^2$ provided in Eq.~\eqref{BIshear}:
\begin{equation}\label{BIshearp}     
    \dot{\sigma}+3H\sigma=0. 
\end{equation}     
Its integration further leads to
\begin{equation}\label{BIsigma2}  
\sigma^2 = \sigma_0^2 a^{-6}.
\end{equation}
Using equations \eqref{sources} and \eqref{BIsigma2} into \eqref{hubble0}, we obtain
\begin{equation}\label{themodel} 
    \frac{H^2}{H_0^2}  =\Omega_{\rm \sigma0}a^{-6}+\Omega_{\text{r}0}a^{-4}+\Omega_{\text{m}0}a^{-3}+\Omega_{\rm de0}a^{-3(1+w_{\rm de0})},
\end{equation}  
where $\Omega_{\sigma 0}=\frac{\sigma_0^2 }{3H_0^2}$, $\Omega_{\rm r0}=\frac{8\pi G}{3H_0^2}\rho_{\rm r0}$, $\Omega_{\rm m0}=\frac{8\pi G}{3H_0^2}\rho_{\rm m0}$ and $\Omega_{\rm de0}=\frac{8\pi G}{3H_0^2}\rho_{\rm de0}$, respectively denote the expansion anisotropy, radiation, matter, and DE density parameters, satisfying $\Omega_{\rm \sigma0} + \Omega_{\text{r}0}+ \Omega_{\text{m}0}+\Omega_{\rm de0}=1$. Here, the expansion anisotropy parameter $\Omega_{\rm \sigma0}$ is purely a geometric term that is non-negative. 

Furthermore, the generalized Friedmann equation \eqref{themodel} depicts the spatially flat and homogeneous but probably non-isotropic Universe (because of the presence of the expansion anisotropy). We denote this model by $w$CDM+$\Omega_{\rm \sigma0}$ and the set of free baseline parameters for this model is given by 
$$\mathcal{P}_{w\rm CDM+\Omega_{\rm \sigma0}}= \left\{ \omega_{\rm b}, \, \omega_{\rm c}, \, H_0, \, \Omega_{\sigma0 }, \, w_{\rm de0}  \right\}.$$
Here $\omega_{\rm b}=\Omega_{\rm b} h^2$ and $\omega_{\rm c}=\Omega_{\rm c}h^2$ are physical density parameters of baryons and CDM, respectively, in the present Universe, $H_0$ is the Hubble constant and $\Omega_{\sigma0 }$ is the expansion anisotropy parameter.

In the absence of expansion anisotropy, equation \eqref{themodel} reduces to 
\begin{equation}\label{themodel1} 
    \frac{H^2}{H_0^2}  =\Omega_{\text{r}0}a^{-4}+\Omega_{\text{m}0}a^{-3}+\Omega_{\rm de0}a^{-3(1+w_{\rm de0})}.   
\end{equation}

It is an isotropic extension of $\Lambda$CDM cosmology. We denote this model by $w$CDM and the set of free baseline parameters for this model is given by  $$\mathcal{P}_{w\rm CDM}= \left\{ \omega_{\rm b}, \, \omega_{\rm c}, \, H_0, \, w_{\rm de0} \right\}.$$
 We follow the standard neutrino scheme with a normal hierarchy which has three species of neutrino, out of which two are massless and one is massive neutrino with the standard number of effective neutrino species, $N_{\rm eff}=3.046$ and minimum allowed mass, $m_{\nu}=0.06\,\rm eV$.

\section{Data and methodology}\label{sec3}
The following data sets have been used in this work: \\\\

\noindent\textbf{Baryon Acoustic Oscillation (BAO)}: Updates on BAO measurements utilizing galaxies, quasars, and Lyman-$\alpha$ (Ly$\alpha$) for completed experiments are provided by the Sloan Digital Sky Survey (SDSS)~\cite{eBOSS:2020yzd}. These experiments involve the compilation of data from BOSS, eBOSS, SDSS, and SDSS-II, making accessible, as indicated in Table \ref{tab:BAO_measurements}, independent BAO measurements of angular-diameter distances and Hubble distances relative to the sound horizon from eight separate samples. 

\begin{table}[b!]
\caption{\rm Clustering measurements for the BAOs  from Ref.~\cite{eBOSS:2020yzd}.}
 \scalebox{0.9}{
\centering
\begin{tabular}{l|c|c|c|c}
\toprule
\hline
\textbf{Parameter} & $\bm{\,\,z_{\rm eff}\,\,}$  &  $\bm{\,\,D_V(z)/r_{\rm d}\,\,}$ & $\bm{\,\,D_M(z)/r_{\rm d}\,\,}$ & $\bm{\,\,D_H(z)/r_{\rm d}\,\,}$  \\
\hline

\hline
MGS & 0.15 & $4.47 \pm 0.17$ & --- & --- \\

BOSS Galaxy & $0.38$ & --- & $10.23 \pm 0.17$ & $25.00 \pm 0.76$ \\

BOSS Galaxy & $0.51$ & --- & $13.36 \pm 0.21$ & $22.33 \pm 0.58$ \\

eBOSS LRG & $0.70$ & --- & $17.86 \pm 0.33$ & $19.33 \pm 0.53$ \\

eBOSS ELG & $0.85$ & $18.33_{-0.62}^{+0.57}$ & --- & --- \\

eBOSS Quasar & $1.48$ & --- & $30.69 \pm 0.80$ & $13.26 \pm 0.55$ \\

Ly$\alpha$-Ly$\alpha$ & $2.33$ & --- & $37.6 \pm 1.9$ & $8.93 \pm 0.28$ \\

Ly$\alpha$-Quasar & $2.33$ & --- &$37.3 \pm 1.7$ & $9.08 \pm 0.34$ \\
\hline
\bottomrule
\hline
\end{tabular} }
\label{tab:BAO_measurements}
\end{table}

At the drag redshift ($z_{\rm d}$), the comoving size of the sound horizon ($r_{\rm s}$) is $r_{\rm d}$, and it is given by
\begin{equation}{
\label{comoving size}
r_{\rm d}=r_{\rm s}(z_{\rm d})=\int_{z_{\rm d}}^\infty \frac{c_{\rm s}\text{d}z}{H(z)},} 
\end{equation}
where $c_{\rm s}=\frac{c}{\sqrt{3(1+\mathcal{R})}}$ is the sound speed of the baryon-photon fluid; $\mathcal{R}=\frac{3\Omega_{\rm b0}}{4\Omega_{\rm \gamma 0}(1+z)}$ with $\Omega_{\rm b0}=0.022h^{-2}$ being the present-day physical density of baryons and $\Omega_{\gamma 0}=2.469\times 10^{-5}h^{-2}$ being the present-day physical density of photons ~\cite{Cooke:2016rky,Bennett:2020zkv}.

Direct constraints on the values $D_H(z)/r_{\rm d}$ and $D_{M}(z)/r_{\rm d}$ are provided by the BAO measurements. We calculate the Hubble distance as follows:
\begin{equation}
 D_H(z) = \frac{c}{H(z)}.
\end{equation}
For flat cosmology, we calculate the comoving angular diameter distance ($D_{M}(z)$)  as follows:
\begin{equation}
    D_{M}(z)= {c \over H_0}\int_0^z \text{d}z' {H_0 \over H(z')}.
\label{eqn:dcomove}
\end{equation}

The spherically averaged distance ($D_V(z)$) is defined as follows:
\begin{equation}
 D_V(z) \equiv \left[z D^2_M(z) D_H(z)\right]^{1/3}.
\end{equation}

We  have opted $d_1$ for $D_{V}(z)/r_{\rm d}$, $d_2$ for $D_{M}(z)/r_{\rm d}$, and $d_3$ for $D_{H}(z)/r_{\rm d}$. Then, the chi-squared function for each measurement in Table \ref{tab:BAO_measurements} is written as follows: 
\begin{eqnarray}
\chi^2_{\rm B_1} =&\displaystyle\sum_{i=1}^{2}\left(\frac{d_1^{\text{obs}}(z_i)-d_1^{\text{th}}(z_i)}{\sigma_{d_1^{\text{obs}}(z_i)}}\right)^2,\nonumber \\
\chi^2_{\rm B_2}=&\displaystyle\sum_{j=1}^{6}\left(\frac{d_2^{\text{obs}}(z_j)-d_2^{\text{th}}(z_j)}{\sigma_{d_2^{\text{obs}}(z_j)}}\right)^2,\nonumber \\
\chi^2_{\rm B_3}=&\displaystyle\sum_{j=1}^{6}\left(\frac{d_3^{\text{obs}}(z_j)-d_3^{\text{th}}(z_j)}{\sigma_{d_3^{\text{obs}}(z_j)}}\right)^2.
\end{eqnarray}

Here, $d^{\rm obs}$ denotes the observed distance value as shown in Table \ref{tab:BAO_measurements}, whereas $d^{\rm th}$ denotes the theoretical value computed for the models under consideration. The effective redshifts for MGS and eBOSS ELG samples in the $D_{V}(z)/r_{\rm d}$ scenario are denoted as $z_i$ ($i=1,2$). For $D_{M}(z)/r_{\rm d}$ and $D_{H}(z)/r_{\rm d}$, the effective redshifts for the six measurements are labeled as $z_j$ ($j=1,2,3,4,5,6$) which correspond to BOSS Galaxy, eBOSS Galaxy, eBOSS LRG, eBOSS Quasar, Ly$\alpha$-Ly$\alpha$, and Ly$\alpha$-Quasar, respectively.

For the BAO measurements, the cumulative chi-squared expression, denoted as $\chi^2_{\rm BAO}$, is therefore formulated as follows:
\begin{equation}
\chi^2_{\rm BAO}=\chi^2_{\rm B_1} +\chi^2_{\rm B_2}+\chi^2_{\rm B_3}.
\end{equation}

\noindent\textbf{Cosmic Chronometer (CC)}: 
The Cosmic Chronometer (CC) uses a reliable technique to monitor the Universe's historical expansion through the analysis of $H(z)$ measurements. 33 $H(z)$ measurements from CC have been compiled, covering redshift values from 0.07 to 1.965. Table \ref{tab:CC} provides a thorough description of these measurements and the corresponding references. The idea behind these measurements was first presented in \cite{Jimenez:2001gg}, whereby a fundamental connection was established between the Hubble parameter $H(z)$, redshift $z$, and cosmic time $t$:
\begin{equation}
    H(z)= -\frac{1}{1+z}\frac{\text{d}z}{\text{d}t}.
\end{equation}

\begin{table}[t!]
\caption{\rm Compilation of CC measurements of $H(z)$.}

\begin{center}
\begin{tabular}{lllr}
\multicolumn{4}{c}{{}}\\
\hline \toprule\hline
$z$ & $H(z)$ & $\sigma_{H(z)}$ & Ref.\\
\hline
0.07 & 69.0 & 19.6 &  \cite{Zhang:2012mp}\\
0.09 & 69 & 12 & \cite{Simon:2004tf}\\
0.12 & 68.6 & 26.2 & \cite{Zhang:2012mp}\\
0.17 & 83 & 8 & \cite{Simon:2004tf}\\
0.179 & 75 & 4 &  \cite{Moresco:2012jh}\\
0.199 & 75 & 5 &  \cite{Moresco:2012jh}\\
0.20 & 72.9 & 29.6 &  \cite{Zhang:2012mp}\\
0.27 & 77 & 14 & \cite{Simon:2004tf}\\
0.28 & 88.8 & 36.6 &  \cite{Zhang:2012mp}\\
0.352 & 83 & 14 & \cite{Moresco:2012jh}\\
0.38 & 83 & 13.5 &  \cite{Moresco:2016mzx}\\
0.4 & 95 & 17 & \cite{Simon:2004tf}\\
0.4004 & 77 & 10.2 & \cite{Moresco:2016mzx}\\
0.425 & 87.1 & 11.2 &  \cite{Moresco:2016mzx}\\
0.445 & 92.8 & 12.9 &  \cite{Moresco:2016mzx}\\
0.47 & 89.0 & 49.6 &  \cite{Ratsimbazafy:2017vga}\\
0.4783 & 80.9 & 9 &  \cite{Moresco:2016mzx}\\
0.48 & 97 & 62 & \cite{Stern:2009ep}\\
0.593 & 104 & 13 &  \cite{Moresco:2012jh}\\
0.68 & 92 & 8 &  \cite{Moresco:2012jh}\\
0.75 & 98.8 & 33.6 &  \cite{Borghi:2021rft}\\
0.781 & 105 & 12 &  \cite{Moresco:2012jh}\\
0.8 & 113.1 &  15.1  &  \cite{Jiao:2022aep}     \\
0.875 & 125 & 17 &  \cite{Moresco:2012jh}\\
0.88 & 90 & 40 &  \cite{Stern:2009ep}\\
0.9 &  117 &  23 &  \cite{Simon:2004tf}\\
1.037 & 154 & 20 &  \cite{Moresco:2012jh}\\
1.3 & 168 & 17 &  \cite{Simon:2004tf}\\
1.363 & 160 & 33.6 &  \cite{Moresco:2015cya}\\
1.43 & 177 & 18 &  \cite{Simon:2004tf}\\
1.53 & 140 & 14 &  \cite{Simon:2004tf}\\
1.75 & 202 & 40 &  \cite{Simon:2004tf}\\
1.965 & 186.5 & 50.4 & \cite{Moresco:2015cya}\\
\hline 
\bottomrule\hline
\end{tabular}
\label{tab:CC}

\end{center}
\end{table}

We formulate the chi-squared function $\chi^2_{\rm CC}$ associated with these measurements as follows:
\begin{equation}
\chi^2_{\rm CC} = \sum_{i=1}^{33} \frac{[H^{\text{obs}}(z_i)-H^{\text{th}}(z_i)]^2}{\sigma^2_{H^{\text{obs}}(z_i)}},
\end{equation}
where $H^{\text{obs}}(z_i)$ signifies the value of the observed Hubble parameter, along with its corresponding standard deviation $\sigma^2_{H^{\text{obs}}(z_i)}$ from the table provided. The equivalent $H^\text{th}(z_i)$ represents the Hubble parameter value in theory obtained from the considered model.\\

\noindent\textbf{Big Bang Nucleosynthesis (BBN)}: 
An independent method of inferring the density of baryons is to examine BBN, which serves as a probe of the early Universe's dynamics. With the help of BBN, we can look at the limitations regarding the anisotropic extensions of the conventional cosmological model \cite{Kneller:2004jz,Steigman:2007xt,Barrow:2018yyg}. For all of our calculations, we use an updated estimate of the physical baryon density, $\omega_b$ (where $\omega_b \equiv \Omega_bh^2$), which comes from Big Bang Nucleosynthesis (BBN) and has a value of $0.02233\pm0.00036$. This computation makes use of revised data obtained from experimental nuclear physics at the INFN Laboratori Nazionali del Gran Sasso in Italy, at the Laboratory for Underground Nuclear Astrophysics (LUNA) ~\cite{Mossa:2020gjc}.\\

\noindent\textbf{Pantheon Plus and SH0ES}:
Historically, Type Ia supernovae (SNe Ia) have played a crucial role in developing the conventional model of the Universe. These supernovae yield useful distance moduli measurements, constraining the late-time expansion rate or the uncalibrated luminosity distance $H_0d_L(z)$. For a supernova at redshift $z$, the theoretical apparent magnitude $m_B$ is determined by the following equation:
\begin{eqnarray}
\label{distance_modulus}
m_B = 5 \log_{10} \left[ \frac{d_L(z)}{1\rm Mpc} \right] + 25 + M_B,
\end{eqnarray}
where $M_B$ stands for the absolute magnitude. Consequently, the distance modulus ($\mu(z)$) can be written as $\mu(z) = m_{B} - M_{B}$. In a flat cosmology, the luminosity distance is determined as follows: 
\begin{equation}
\label{eq:dl}
d_L(z) = (1+z)\int_0^{z}\frac{dz^\prime}{H(z^\prime)}.
\end{equation}

In the present study, we use SNe Ia distance modulus measurements from the Pantheon+ sample \cite{Brout:2022vx}. We name the collection PP, which comprises 1701 light curves that correspond to 1550 unique supernovae Ia within the $z \in [0.01, 2.26]$ redshift range. We also include constraints on $H_0$ and $M_B$ by including SH0ES Cepheid host distance anchors \cite{Brout:2022vx} into our analysis; this dataset is referred to as PPSH0ES.

We utilize PP data to minimize the equation that follows to constrain the cosmological parameters:
\begin{equation}
\label{eq:likelihood}
\chi^2_{\rm PP} = \Delta D^T~C_{\rm stat+syst}^{-1}~\Delta D ,
\end{equation}
where $D$ represents the vector of 1701 SN distance modulus residuals calculated as 
\begin{equation}
\label{eq:dmu}\Delta D_i = \mu_i - \mu_{{\rm model}}(z_i) ,
\end{equation}
 while comparing the observed supernova distance ($\mu_i$) with the predicted model distance ($\mu_{{\rm model}}(z_i)$) by utilising the measured supernova/host redshift. Here, $C_{\rm stat+syst} = C_{\rm stat} + C_{\rm syst}$, with $C_{\rm stat}$ and $C_{\rm syst}$ denote the statistical and the systematic covariance matrices respectively.

While SNe analysis alone suffers from degeneracy between the parameters $M_B$ and $H_0$, we overcome this restriction by adding the recently reported SH0ES Cepheid host distance anchors (R22) into the likelihood. With this change, we can confine $M_B$ as well as $H_0$. The following adjustments are made to the SN distance residuals after including SH0ES Cepheid host distances:
\begin{equation}
\label{eq:dmuprime}
\Delta D^\prime_i=
        \begin{cases}
            \mu_i - \mu_i^{{\rm Cepheid}} & i \in \text{Cepheid hosts} \\
            \mu_i - \mu_{{\rm model}}(z_i) &\text{otherwise} ,
        \end{cases}
    \end{equation}
where $\mu_i^{{\rm Cepheid}}$ stands for the Cepheid-calibrated host-galaxy distance from SH0ES. After incorporating this data with the SH0ES Cepheid host-distance covariance matrix ($C^{\rm Cepheid}_{\rm stat+syst}$) from R22, the likelihood function is modified as follows:
\[ \chi^2_{\rm PPSH0ES} = \Delta D{^\prime}^T~(C^{\rm SN}_{\rm stat+syst}+C_{\rm stat+syst}^{\rm Cepheid})^{-1}~\Delta D^\prime, \]
where $C^{\rm SN}_{\rm stat+syst}$ represents the SN covariance. Further information is available in the reference \cite{Brout:2022vx}.\\


 The baseline free parameters of the model $w$CDM+$\Omega_{\sigma0 }$ and $w$CDM are as discussed in section \ref{sec2}. 
 We have assumed three neutrino species, approximated as two massless states and a single massive neutrino of mass $m_{\nu}=0.06\,\rm eV$. We have used uniform priors: $\omega_{\rm b}\in[0.01,0.03]$, $\omega_{\rm c}\in[0.05,0.25]$, $\,H_0\in[60,80]$, $\Omega_{\sigma0 }\in[0,0.001]$, $w_{\rm de0} \in [-2,0]$ and the publicly available Boltzmann code \texttt{CLASS}~\cite{Blas:2011rf} with the parameter inference code  \texttt{Monte Python}~\cite{Audren:2012wb} to obtain correlated Monte Carlo Markov Chain (MCMC) samples. We analyze the MCMC samples using the python package \texttt{GetDist}\footnote{\href{https://getdist.readthedocs.io/en/latest/intro.html}{https://getdist.readthedocs.io/en/latest/intro.html}}. It is important to note here, we use the above-mentioned data in two combinations: BAO+CC+BBN+PP and BAO+CC+BBN+PPSH0ES, and abbreviated as  $D_1$ and $D_2$, respectively throughout the forthcoming text.

\section{Results and Discussion}\label{sec4}

\begin{figure}
	\centering
	\includegraphics[width=10cm]{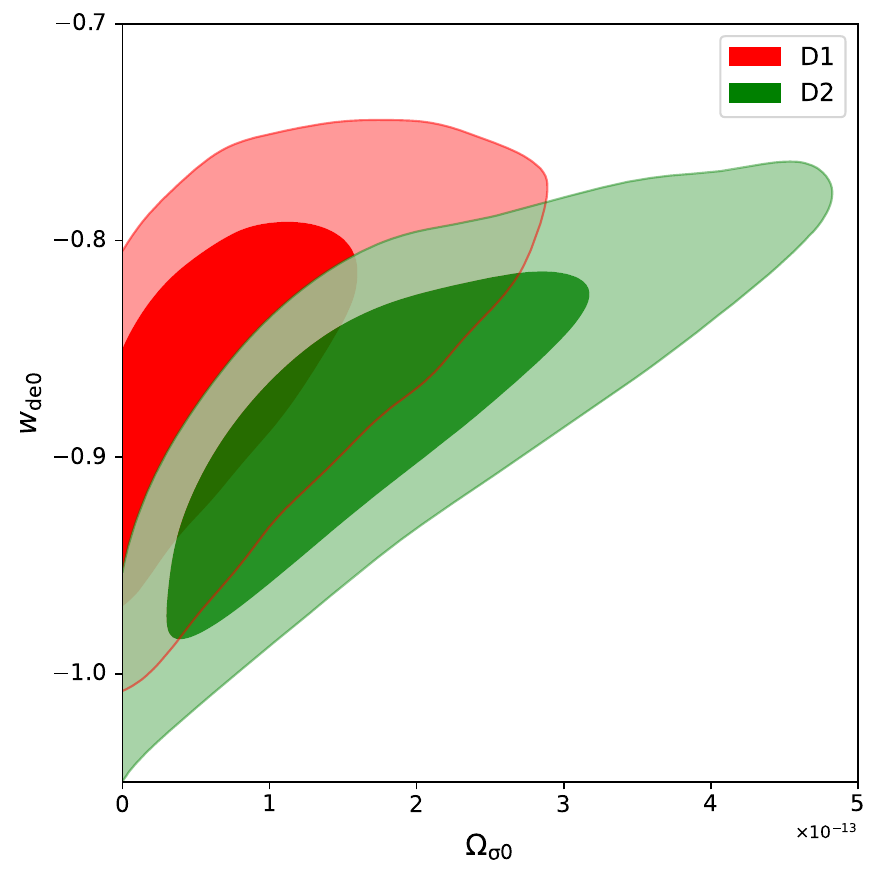}
	\caption{Two-dimensional marginalized confidence regions at 68\% and 95\% CL for $\Omega_{\rm \sigma0}$ and $w_{\rm de0}$ for the $w$CDM+$\Omega_{\rm \sigma0}$ model from $D_1$ and $D_2$ data.}
	\label{fig_desigma}
\end{figure}

In Table \ref{tab:1}, we summarize the constraints on free parameters and some derived parameters of the $w$CDM and $w$CDM+$\Omega_{\rm \sigma0}$ models at $68\%$CL from both combinations: $D_1$ and $D_2$. The second row over each parameter (mentioned in blue color) represents the constraints at $68\%$CL  for $\Lambda$CDM and $\Lambda$CDM+$\Omega_{\rm \sigma0}$ models from both data sets. The upper bounds on the anisotropic expansion parameter, $\Omega_{\rm \sigma0}$ at $95\%$ CL are of order $10^{-13}$ for $w$CDM+$\Omega_{\rm \sigma0}$ model with both $D_1$ and $D_2$. The upper bounds of $\Omega_{\rm \sigma0}$ for this model read as: $\Omega_{\rm \sigma0}<2.3\times 10^{-13}$  and $\Omega_{\rm \sigma0}<4\times 10^{-13}$ from $D_1$ and $D_2$, respectively. On the other hand, the upper bounds on the anisotropic expansion parameter $\Omega_{\rm \sigma0}$ for $\Lambda$CDM+$\Omega_{\rm \sigma0}$ model are: $\Omega_{\rm \sigma0}<2.0\times 10^{-14}$ and  $\Omega_{\rm \sigma0}<6.5\times 10^{-14}$ from $D_1$ and $D_2$, respectively. Thus, we see that the $w$CDM+$\Omega_{\rm \sigma0}$ model provides weaker upper bounds on the anisotropy parameter by one order of magnitude as compared to $\Lambda$CDM+$\Omega_{\rm \sigma0}$ model with both the data combinations. We observed that the anisotropic parameter $\Omega_{\rm \sigma0}$ finds a strong positive correlation with the DE EoS parameter from $D_1$ and $D_2$. This means that a larger amount of anisotropy in the Universe results in a larger value of the DE EoS, see Figure \ref{fig_desigma}.

\begin{table*}[ht]
  \caption{Constraints at 68\% CL on  the free and some derived parameters of the $w$CDM and $w$CDM+$\Omega_{\rm \sigma0}$ models from $D_1$ and $D_2$ data. The upper bounds on $\Omega_{\rm \sigma0}$ are displayed at 95\% CL. The Hubble constant $H_0$ is measured in the unit of $\rm km\, s^{-1} \rm Mpc^{-1}$. The entries in the second row over each parameter (in blue color) represent the constraints on $\Lambda$CDM and $\Lambda$CDM+$\Omega_{\rm \sigma0}$ model parameters.}
  \label{tab:1}
	\scalebox{1.1}{
	\begin{centering}
	  \begin{tabular}{lcccc}
  	\hline\toprule \hline
    \multicolumn{1}{l}{Data set} & \multicolumn{2}{c}{$D_1$} & \multicolumn{2}{c}{$D_2$} \\  \hline
    Model &$w$CDM & $w$CDM+$\Omega_{\rm \sigma0}$ & $w$CDM  & $w$CDM+$\Omega_{\rm \sigma0}$    \\
    & \textcolor{blue}{$\Lambda$CDM} & \textcolor{blue}{ $\Lambda$CDM+$\Omega_{\rm \sigma0}$} & \textcolor{blue}{$\Lambda$CDM} & \textcolor{blue}{$\Lambda$CDM+$\Omega_{\rm \sigma0}$}    \\ \hline   
      
 $10^{-2}\omega_{\rm b }$ & $2.237^{+0.035}_{-0.035} $ &  $2.231^{+0.036}_{-0.036}$ & $2.248^{+0.035}_{-0.035}  $ &   $2.231^{+0.036}_{-0.036}$ \\
   & \textcolor{blue}{$2.233^{+0.035}_{-0.035}$} & \textcolor{blue}{$2.230^{+0.036}_{-0.036}$} & \textcolor{blue}{$2.258 ^{+0.035}_{-0.035} $} &  \textcolor{blue}{$2.232^{+0.036}_{-0.036} $} \\ 
 \hline
 \vspace{0.1cm}
 
 $\Omega_{\rm \sigma0 }$ & 0 & $<2.3\times 10^{-13}\;(95\%\; \text{CL})$  & 0 & $<4\times 10^{-13}\;(95\%\; \text{CL})$    \\
   &   \textcolor{blue}{0} & \textcolor{blue}{$<2.0\times 10^{-14}\;(95\%\; \text{CL})$ } & \textcolor{blue}{0} &\textcolor{blue}{ $<6.5\times 10^{-14}\;(95\%\; \text{CL})$}  
   \\
\hline
 \vspace{0.1cm}
 
 $w_{\rm de0}$  &    $-0.930^{+0.045}_{-0.045}  $  &  $-0.853^{+0.054}_{-0.048}   $    & $ -1.062^{+0.041}_{-0.041}   $ & $ -0.881^{+0.053}_{-0.053}   $\\
 
 &   \textcolor{blue}{$-1$} & \textcolor{blue}{$-1$} & \textcolor{blue}{$-1$} &\textcolor{blue}{ $-1$}  
   \\
 \hline

 \vspace{0.10cm} 
$H_{\rm 0}$  &    $66.4^{+1.3}_{-1.3} $  & $68.4^{+1.6}_{-1.6}  $    & $ 71.51^{+0.83}_{-0.83} $ & $ 72.24^{+0.82}_{-0.82}  $\\
   & \textcolor{blue}{$68.02^{+0.84}_{-0.84}$} & \textcolor{blue}{$70.1^{+1.2}_{-1.5} $ }  & \textcolor{blue}{$ 70.79^{+0.69}_{-0.69}$} & \textcolor{blue}{$72.67^{+0.85}_{-0.85}$}\\
\hline
 \vspace{0.1cm}
 
$M_{B}$  & $-19.461^{+0.040}_{-0.040}$ & $-19.395^{+0.047}_{-0.047} $ & $-19.317^{+0.023}_{-0.023}$ & $-19.282^{+0.023}_{-0.023}$ \\
 &  \textcolor{blue}{$-19.418^{+0.028}_{-0.028}$} & \textcolor{blue}{ $-19.356^{+0.039}_{-0.044}$ }&\textcolor{blue}{$-19.330^{+0.020}_{-0.022}$} & \textcolor{blue}{$-19.283^{+0.024}_{-0.024}$} \\
\hline
 \vspace{0.10cm} 

 $\Omega_{\rm m0 }$ &   $0.258^{+0.015}_{-0.015}   $   & $0.230^{+0.018}_{-0.018}  $  & $0.287^{+0.012}_{-0.012}  $ & $0.233^{+0.018}_{-0.018} $     \\
   &  \textcolor{blue}{$0.318^{+0.011}_{-0.011}$}   & \textcolor{blue}{$0.309^{+0.012}_{-0.012}$} & \textcolor{blue}{$0.324^{+0.011}_{-0.011}$} & \textcolor{blue}{$0.302^{+0.012}_{-0.012}$}    \\
\hline
 \vspace{0.10cm} 
 $z_{\rm d}$ &   $1059.3^{+1.1}_{-1.1} $   & $1058.7^{+1.1}_{-1.1}  $ & $1061.99^{+0.88}_{-0.88}$ & $1059.8^{+1.1}_{-1.1} $     \\
    &  \textcolor{blue}{ $1060.09^{+0.95}_{-0.95}$ }  & \textcolor{blue}{$1060.37^{+0.98}_{-0.98}$ } & \textcolor{blue}{$1061.75^{+0.85}_{-0.85}$} & \textcolor{blue}{$1060.99^{+0.90}_{-0.90}$}    \\
 \hline
 \vspace{0.10cm} 
 $r_{\rm d}$ &   $149.0^{+2.7}_{-2.7}$   & $144.4^{+3.1}_{-3.1}$ & $140.7^{+1.8}_{-1.8} $ &  $137.7^{+1.9}_{-1.9}  $   \\ 
    &  \textcolor{blue}{$146.1^{+1.8}_{-1.8}$}  & \textcolor{blue}{$142.0^{+2.7}_{-2.7}$} &  \textcolor{blue}{$142.1^{+1.5}_{-1.5}$} & \textcolor{blue}{$137.7^{+1.9}_{-1.9} $} \\ \hline
 \vspace{0.10cm} 
 $\chi^2_{\rm min}$ &   $1436.2$   &  $1434.08$  & $1332.1$ & $1319.9$ \\
     &  \textcolor{blue}{$1438.62$}   & \textcolor{blue}{$1438.58$}  & \textcolor{blue}{$1334.4$} & \textcolor{blue}{$1322.94$} \\

\hline\bottomrule\hline

\end{tabular}
\end{centering}}
\end{table*}

\begin{figure}
	\centering
	\includegraphics[width=10cm]{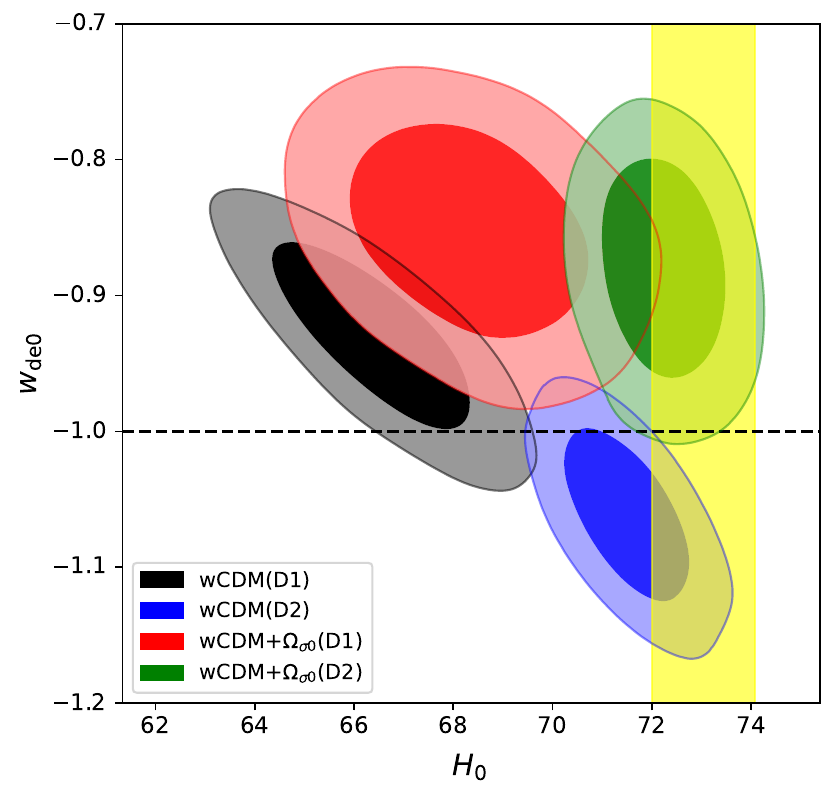}
	\caption{Two-dimensional marginalized confidence regions at 68\% and 95\% CL of $H_0$ and $w_{\rm de0}$ for the $w$CDM and $w$CDM+$\Omega_{\rm \sigma0}$ models from $D_1$ and $D_2$ data. The vertical yellow band represents $H_{\rm 0}^{\rm R22}=73.04\pm1.04~{\rm km\, s^{-1}\, Mpc^{-1}}$ from  SH0ES measurement. The horizontal dotted black line represents $w_{\rm de0 } = -1$.}
	\label{fig_wde0_H0}
\end{figure}

 For the $w$CDM+$\Omega_{\rm \sigma0}$ model, it can be observed from Table \ref{tab:1} that the DE EoS parameter does not prefer the cosmological constant form of the DE (having $w_{\rm de0}=-1$) with $D_1$ and $D_2$. Also, at $95\%$ CL, the DE EoS parameter ($w_{\rm de0} = -0.853^{+0.096}_{-0.100}$ for $D_1$ and $w_{\rm de0} = -0.88^{+0.10}_{-0.11}$ for $D_2$) does not prefer the cosmological constant form of the DE. For this model, the mean values of the DE EoS parameter (at $68\%$ CL as well as $95\%$ CL) lie in the quintessence region $(w_{\rm de0}> -1)$, pointing towards the quintessence behavior of the DE. Also, for the $w$CDM model, the mean value of $w_{\rm de0}$  at $68 \%$ CL,  as shown in Table \ref{tab:1}, lies in the quintessence region with $D_1$ whereas, it touches the phantom barrier at $95 \%$ CL with $D_1$ ($w_{\rm de0} = -0.930^{+0.086}_{-0.090}$). Again, for the same model with $D_2$, the mean value of $w_{\rm de0}$ lies in the phantom region $(w_{\rm de0}< -1)$ at $68 \%$ CL as well as $95 \%$ CL with $w_{\rm de0} = -1.062^{+0.078}_{-0.082}$ at $95 \%$ CL. Also, we observe from Figure \ref{fig_wde0_H0} that $w_{\rm de0}$ finds a strong negative correlation with $H_0$ in the case of $w$CDM model whereas a slight negative correlation in the case of $w$CDM+$\Omega_{\rm \sigma0}$ model with both $D_1$ and $D_2$ combinations.

\begin{figure*}
	\centering
	\includegraphics[width=8.5cm]{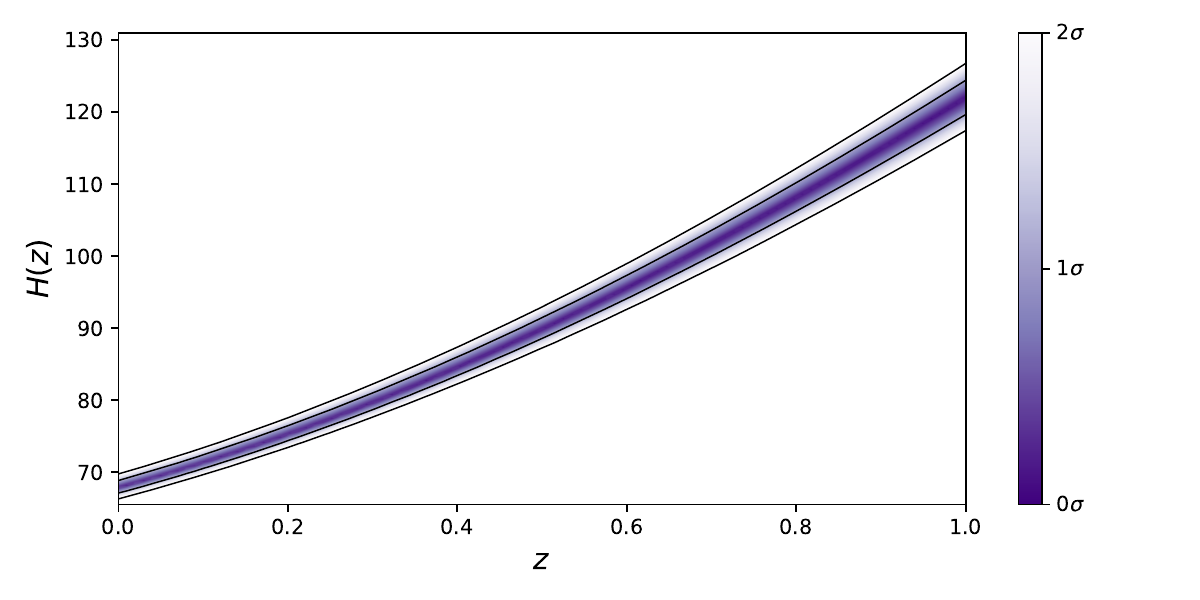}
 \includegraphics[width=8.5cm]{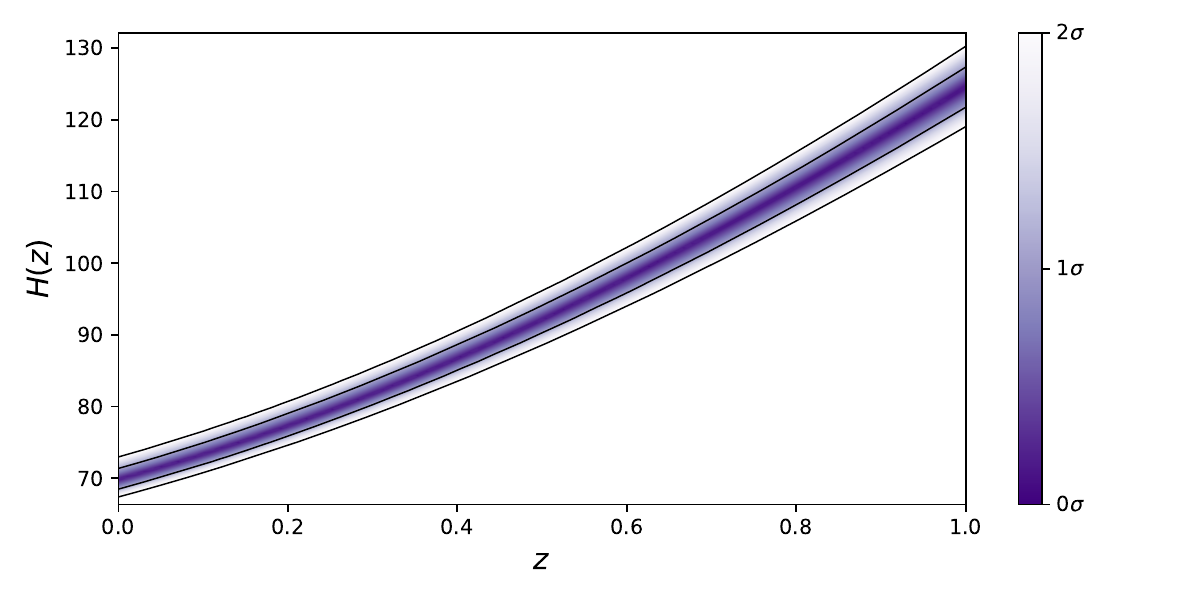}
	\caption{Reconstruction plots of $H(z)$ within 1$\sigma$ and 2$\sigma$ for the $\Lambda$CDM (left) and $\Lambda$CDM+$\Omega_{\rm \sigma0}$ (right) models using the best-fit values from $D_1$ data. }
	\label{fig_hzplots}
\end{figure*}

\begin{figure*}
	\centering
	\includegraphics[width=8cm]{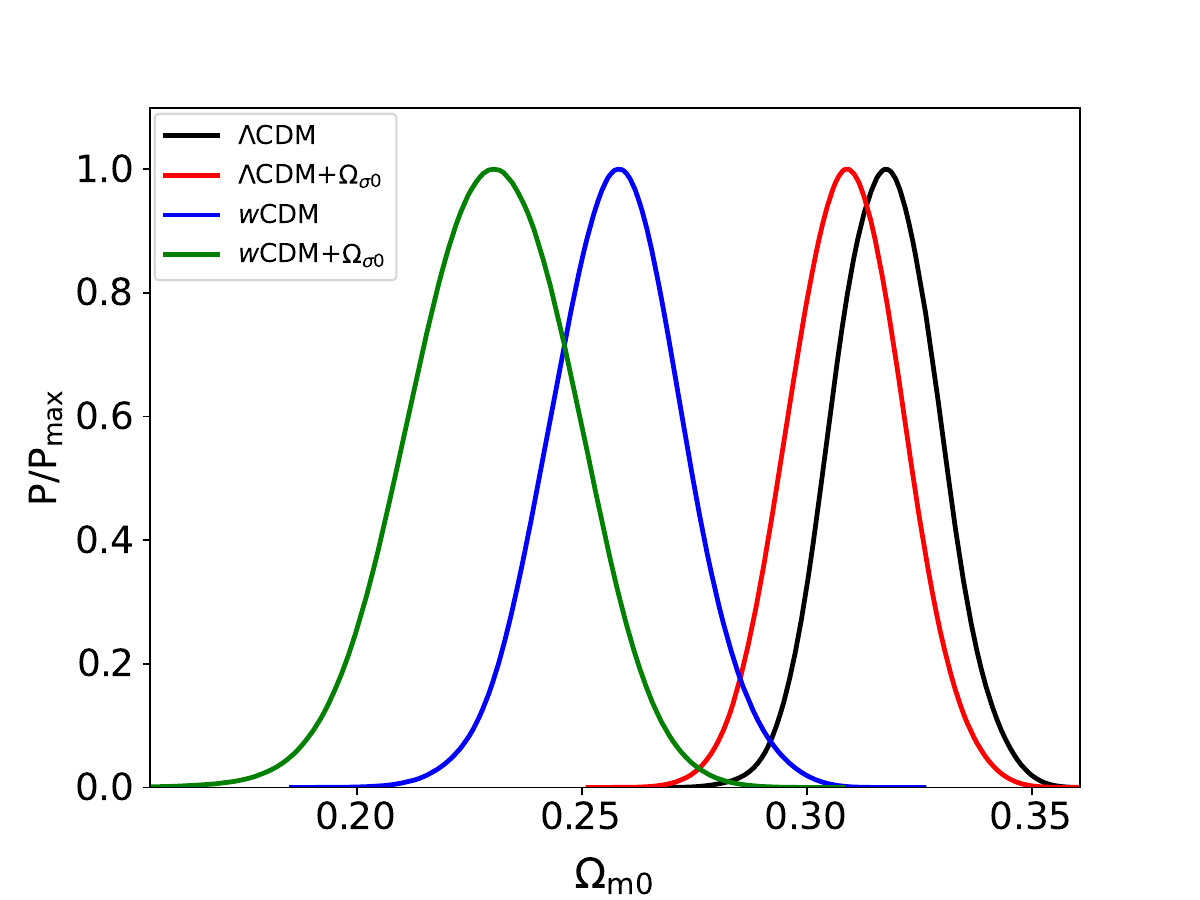}
 \includegraphics[width=8cm]{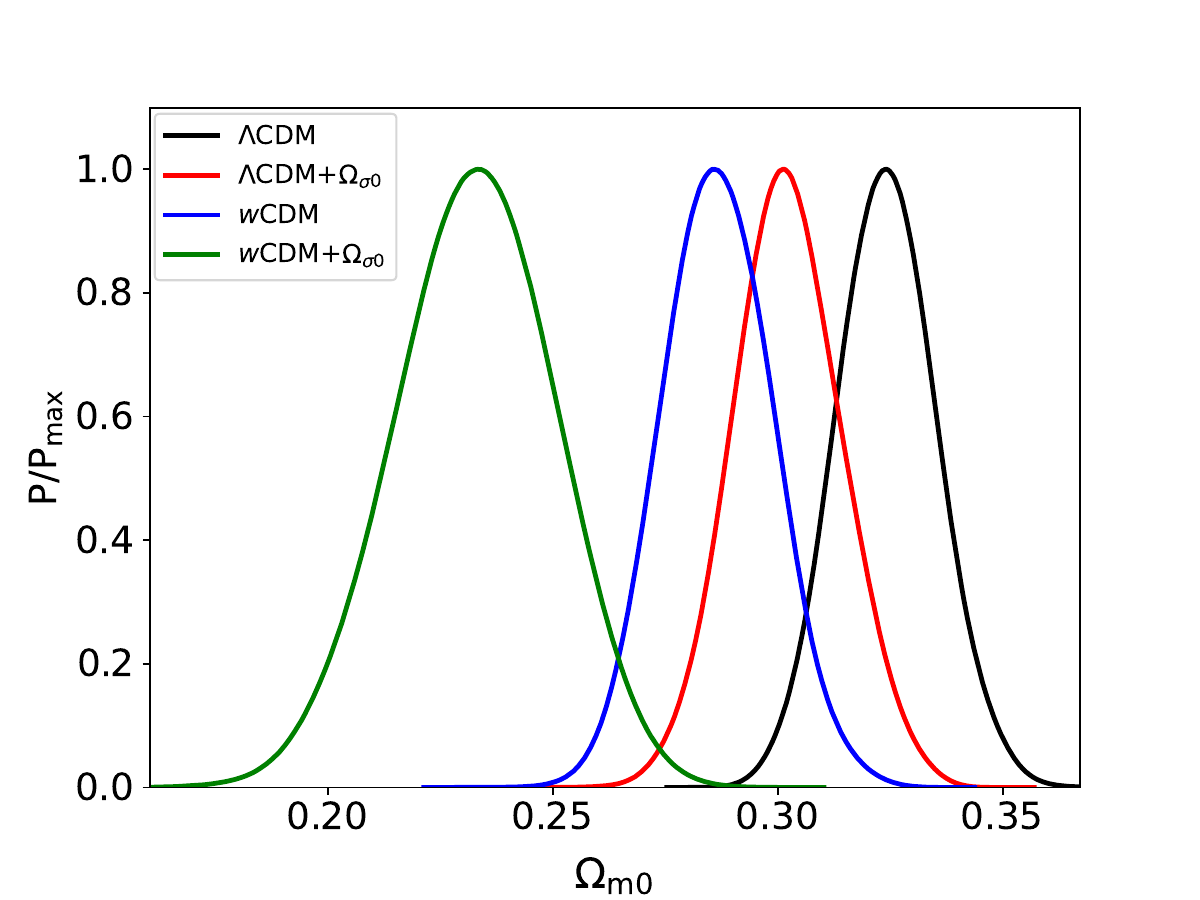}
	\caption{One-dimensional marginalized probability distributions of $\Omega_{\rm m0}$ for the considered models from $D_1$ (left) and $D_2$ (right) combinations. }
	\label{fig_1dplots}
\end{figure*}

\begin{figure*}
	\centering
	\includegraphics[width=17cm]{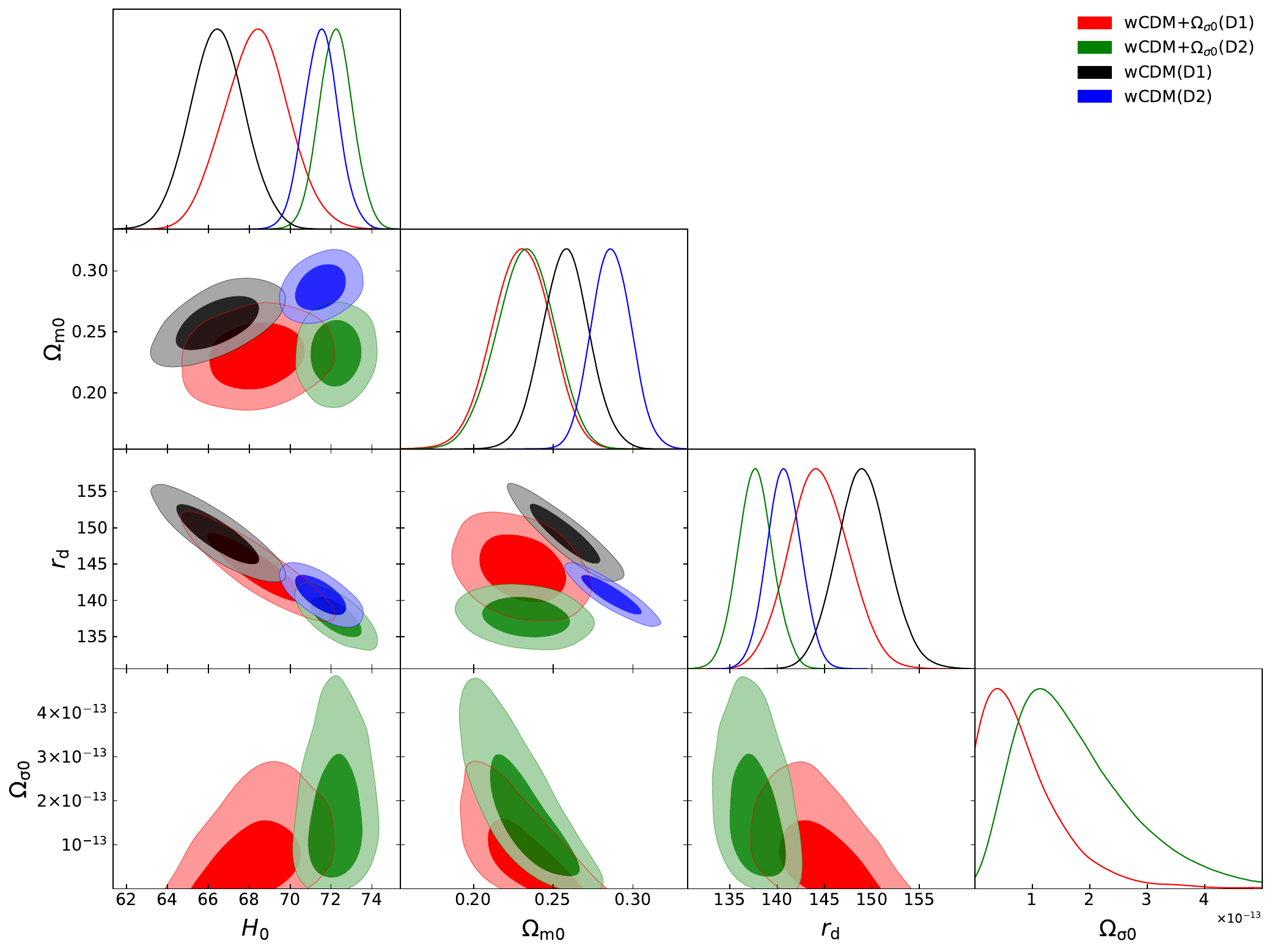}
	\caption{One and two dimensional marginalized confidence regions at  68\% and 95\% CL for some selected parameters  of the $w$CDM and $w$CDM+$\Omega_{\rm \sigma0}$ models from $D_1$ and $D_2$ combinations.} 
	\label{fig_contour}
\end{figure*}

One of the most well-known and intriguing puzzle nowadays in cosmology is a discrepancy between Planck CMB data within the baseline $\Lambda$CDM model and direct local measurements of the current rate of expansion of the Universe (that is, the value of Hubble constant, $H_0$), as discussed in Section \ref{sec:intro}.
Now, we discuss the constraints on the Hubble constant for the considered models and compare the results with the previous work \cite{Yadav:2023yyb}. For combination $D_1$, the constraint on $H_0$ for  the $w$CDM ($\Lambda$CDM) model reads:  $H_{\rm 0}= 66.4 \pm 1.3 \, (68.02 \pm 0.84)~{\rm km\, s^{-1}\, Mpc^{-1}}$, and for the $w$CDM+$\Omega_{\rm \sigma0}$ ($\Lambda$CDM+$\Omega_{\rm \sigma0}$) model, it reads: $H_{\rm 0}= 68.4 \pm 1.6 \, (70.1^{+1.2}_{-1.5})~{\rm km\, s^{-1}\, Mpc^{-1}}$ at 68\% CL. For $D_1$ data, we have obtained lower mean values of $H_0$ for the $w$CDM  and $w$CDM+$\Omega_{\rm \sigma0}$ models as compared to the $\Lambda$CDM and $\Lambda$CDM+$\Omega_{\rm \sigma0}$ models, respectively. 
For combination $D_2$, the constraint on the Hubble constant for the $w$CDM ($\Lambda$CDM) reads:  $H_{\rm 0}= 71.51 \pm 0.83 \, (70.79 \pm 0.69)~{\rm km\, s^{-1}\, Mpc^{-1}}$, and for the $w$CDM+$\Omega_{\rm \sigma0}$ ($\Lambda$CDM+$\Omega_{\rm \sigma0}$) model, it reads: $H_{\rm 0}= 72.24 \pm 0.82 \, (72.67 \pm 0.85)~{\rm km\, s^{-1}\, Mpc^{-1}}$ at 68\% CL. Thus, the $w$CDM model provides higher mean value of $H_0$ as compared to the $\Lambda$CDM model, whereas the $w$CDM+$\Omega_{\rm \sigma0}$ and $\Lambda$CDM+$\Omega_{\rm \sigma0}$ models provide more or less the same mean values with $D_2$. The highest value of the Hubble constant $H_0$ obtained in the present analysis is $72.24 \pm 0.82~{\rm km\, s^{-1}\, Mpc^{-1}}$ from $D_2$ which is consistent with $H_0^{\rm R22}=73.04\pm1.04~{\rm km\, s^{-1}\, Mpc^{-1}}$ from the SH0ES measurement. Quantifying the $H_0$ tension with the SH0ES measurement from $D_1$, we found that there is $3.9\sigma$ tension on $H_0$  in the $w$CDM model and there is $2.4\sigma$ tension on $H_0$ in the $w$CDM+$\Omega_{\rm \sigma0}$ model. Also, for $D_2$, $1.1\sigma$ tension on $H_0$ is observed in the $w$CDM model and $0.6\sigma$ tension is observed on $H_0$ in the $w$CDM+$\Omega_{\rm \sigma0}$ model. So, from $D_1$ the $H_0$ tension is relieved by $1.5\sigma$ in the presence of anisotropy of the order $10^{-13}$ with a constant EoS parameter of the DE. Also, from $D_2$, the $H_0$ tension is relieved by $0.5\sigma$ in the presence of anisotropy of the order $10^{-13}$ with a constant EoS parameter of the DE. See the Figure \ref{fig_wde0_H0} where vertical yellow band represents $H_{\rm 0}^{\rm R22}=73.04\pm1.04~{\rm km\, s^{-1}\, Mpc^{-1}}$ from the SH0ES measurement and constraints on $H_0$ in Table \ref{tab:1}. Also, from Figure \ref{fig_contour} we have observed a positive correlation of $H_0$ with $\Omega_{\rm \sigma0}$ with $D_1$ and $D_2$ both. So, to get a larger value of $H_0$, a larger amount of anisotropy is required. We have shown a reconstruction within 1$\sigma$ and 2$\sigma$ for the function $H(z)$ for the $\Lambda$CDM and $\Lambda$CDM+$\Omega_{\sigma 0}$ models using the best-fit values from $D_1$ analysis to illustrate the impact of $\Omega_{\sigma 0}$ on the expansion rate of the Universe, see Figure \ref{fig_hzplots}. We see that the inclusion of $\Omega_{\sigma 0}$ not only shifts the mean value of $H_0$ but also enlarges the probability range with both data analyses. Thus, the anisotropy parameter (even a low value) has a significant impact on the Hubble parameter, see constraints on $H_0$ in Table \ref{tab:1}.

\begin{figure*}
	\centering
	\includegraphics[width=10cm]{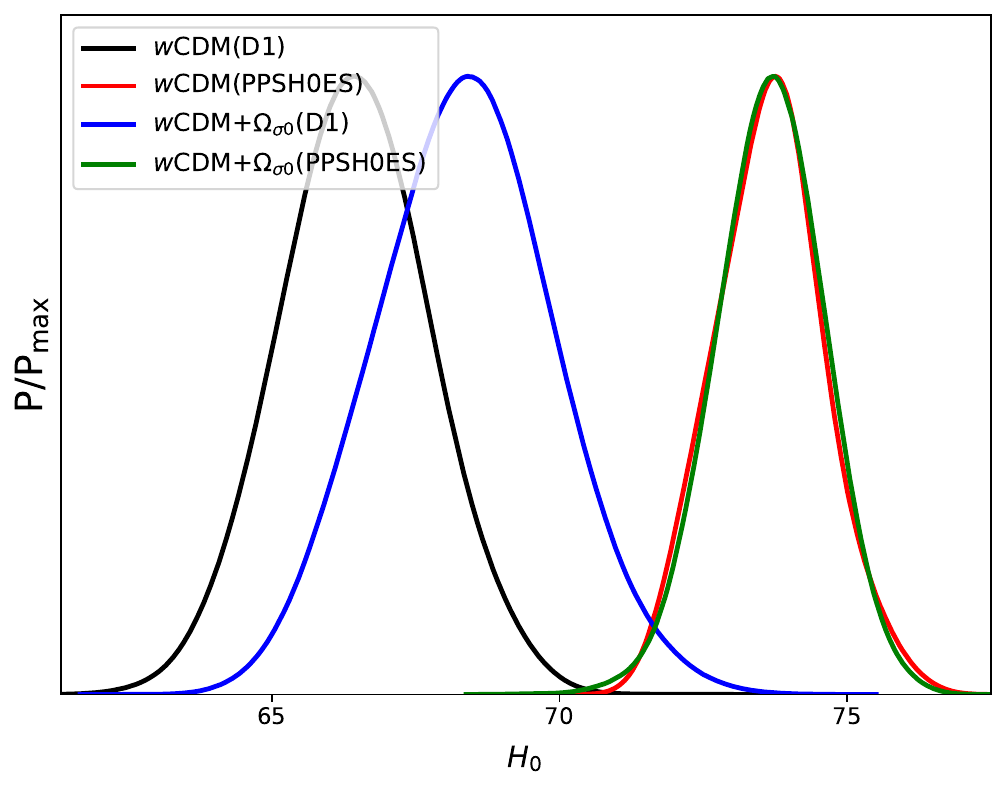}
	\caption{One-dimensional marginalized probability distributions of $H_0$ for the $w$CDM and $w$CDM+$\Omega_{\rm \sigma0}$ models from $D_1$ and PPSH0ES data.} 
	\label{fig_H0}
\end{figure*}
 
 Now, we discuss how the present DE EoS, $w_{\rm de0}$ and $\Omega_{\rm \sigma0}$ affect the present day matter density parameter, $\Omega_{\rm m0}$. Firstly, we observed that the $w$CDM model provides significantly lower mean values of $\Omega_{\rm m0}$ compared to the $\Lambda$CDM model from $D_1$ and $D_2$. Second, adding anisotropy to the $\Lambda$CDM model slightly reduces the mean value of $\Omega_{\rm m0}$ for both combinations. Also, the inclusion of anisotropy into the $w$CDM model significantly reduces the mean values of  $\Omega_{\rm m0}$ for $D_1$ and $D_2$. The constraints on $\Omega_{\rm m0}$ for the $w$CDM model are:  $\Omega_{\rm m0 }=0.258\pm 0.015$ and $\Omega_{\rm m0 }= 0.287\pm 0.012$ from $D_1$ and $D_2$, respectively. The constraints on $\Omega_{\rm m0}$ for the $w$CDM+$\Omega_{\rm \sigma0}$ model are:  $\Omega_{\rm m0 }=0.230\pm 0.018$ from $D_1$ and $\Omega_{\rm m0 }= 0.233\pm 0.018$ from $D_2$. We can observe the deviations in the distribution of $\Omega_{\rm m0}$ for considered models in one-dimensional marginalized probability distribution as shown in Figure \ref{fig_1dplots} (also, see table \ref{tab:1}).  We also observe that the parameters, $\Omega_{\rm \sigma0}$ and $\Omega_{\rm m0 }$ finds a negative correlation with both data combinations, see $\Omega_{\rm \sigma0}-\Omega_{\rm m0 }$ parametric space in Figure \ref{fig_contour}.

The considered $w$CDM and $w$CDM+$\Omega_{\rm \sigma0}$ models provides higher values of $r_{\rm d}$ as compared to the $\Lambda$CDM and $\Lambda$CDM+$\Omega_{\rm \sigma0}$ models with $D_1$. Also, from $D_2$, we obtain the result that the $w$CDM model prefers  a lower value of $r_{\rm d}$ than the $\Lambda$CDM model, but the $\Lambda$CDM+$\Omega_{\rm \sigma0}$ and the $w$CDM+$\Omega_{\rm \sigma0}$ provide equivalent mean values on $r_{\rm d}$ with similar error bars. Further, we notice that adding the anisotropy reduces the value of $r_{\rm d}$ in both cases. We can observe from Figure \ref{fig_contour}, that the parameters $\Omega_{\rm \sigma0}$ and $r_{\rm d}$ are negatively correlated, which means that for a small increment in $\Omega_{\rm \sigma0}$ values, there would be corresponding decrement in $r_{\rm d}$ values.

We have observed that different model parameters are influenced by anisotropy. For instance, we observe a negative correlation between $H_0$ and $w_{\rm de0}$ which can be seen in Figure \ref{fig_wde0_H0}. Note that parameters influence each other while fitting the multi-parameter space simultaneously \cite{Marra:2021fvf,Camarena:2023rsd}. Also, the other parameters of the models under consideration (with constant EoS of DE) are affected by anisotropy in such a way that relieves the $H_0$ tension up to $\sim 1\sigma$. It is important to emphasize that in equation \eqref{themodel}, the scaling $a^{-6}$ has relevance for the early Universe while it becomes negligible in the late Universe. Thus, the anisotropic model under consideration behaves like the Early DE model. For more about Early DE models, we refer the readers to the recent and interesting review \cite{Poulin:2023lkg} and references therein.
\subsection{STATISTICAL MODEL COMPARISON}
The model comparison criteria, play a significant role in evaluating the sustainability of the models. We apply well-known statistical criteria here: the Akaike Information Criteria (AIC) \cite{akaike1974new,burnham2011aic}. The AIC is used to figure out the goodness-of-fit of different models. The model with the lowest AIC is the most preferred by observations. The AIC considers the maximum likelihood function and the number of free parameters of the model. 
It is defined as 
 \begin{equation} \nonumber
 \text{AIC} = -2 \ln  \mathcal{L}_{\rm max} + 2\, \texttt{N} \quad = \chi_{\rm min}^2 + 2\, \texttt{N},
 \end{equation}
where $ \mathcal{L}_{\rm max}$ indicates the maximum likelihood function and  $\texttt{N}$ denotes the total number of free parameters in the model. To compare the derived model with a reference model, the difference of AIC values, i.e.,  $\Delta\text{AIC} =  \text{AIC}_{\rm model}- \text{AIC}_{\rm reference}$ play a crucial role than the AIC values itself. One may say a model is more favored than the other if $\Delta\text{AIC}$ is greater than a threshold value, $\Delta_{\rm threshold}$ \cite{tan2012reliability}.  The thumb rule of AIC states that a universal threshold, $\Delta_{\rm threshold} = 5$ can be used for the model comparison.  In \cite{liddle2007information}, the author mentioned that depending on the minimum threshold AIC difference, one model is preferred over the other. 
\begin{table}[h]\centering
\caption{\label{evidence}{} The difference, $\Delta\text{AIC} = \text{AIC}_{\rm model}- \text{AIC}_{\rm \Lambda CDM}$ of considered models with respect to $\Lambda$CDM model with $D_1$ and $D_2$.}
\begin{tabular}{c c c c c}
\hline \hline
Data set & $\Lambda$CDM+ $\Omega_{\sigma 0}$ & $w$CDM & $w$CDM + $\Omega_{\sigma 0}$ \\ \hline

$D_1$  & $2.38$ & $0$ & $-0.12$ \\
$D_2$ & $-9.5$ & $-0.3$  &$ -10.5$ \\


\hline \hline
\end{tabular}
 \end{table}
 
 The $\Delta \rm AIC$ values of the considered models (with respect to the $\Lambda$CDM) are shown in Table \ref{evidence} for $D_1$ and $D_2$ combinations. We see that from $D_1$, the difference between the models is not statistically significant as the values are less than $\Delta_{\rm threshold}$. From $D_2$, we observe that the $w$CDM model has no statistical preference over the standard model whereas the $\Lambda$CDM+ $\Omega_{\sigma 0}$ and $w$CDM + $\Omega_{\sigma 0}$ show a strong preference over the $\Lambda$CDM model. Observing a preference via the Bayesian model comparison is not surprising as we know that the last term in the AIC acts as a penalty for models with many parameters.  

 In Figure \ref{fig_H0}, we show one-dimensional marginalized probability distributions of $H_0$ for the $w$CDM and $w$CDM+$\Omega_{\rm \sigma0}$ models from $D_1$ and PPSH0ES data. We notice a significant overlapping of $H_0$ values within $w$CDM+$\Omega_{\rm \sigma0}$ model with $D_1$ and PPSH0ES data but that is not the case within the $w$CDM model. So it may be treated as a statistical caveat of the joint analyses with the PPSH0ES data that there exist significant tensions among $D1$ and PPSH0ES datasets within the models under consideration, especially within the $w$CDM model.

\section{Concluding Remarks}
\label{sec5}
In the present work, we have derived CMB independent constraints on two simplest extensions of the $\Lambda$CDM model, namely the $w$CDM and $w$CDM+$\Omega_{\rm \sigma0}$ models from recent data sets including BAO, CC, BBN, PP and PPSH0ES in two combinations: $D_1$ and $D_2$.  We have obtained that the DE EoS parameter favors the quintessence form for the $w$CDM+$\Omega_{\rm \sigma0}$ model from both data combinations at $68\%$ CL as well as $95\%$ CL. For the $w$CDM model, the mean value of $w_{\rm de0}$  at $68 \%$ CL lies in the quintessence region whereas, it touches the phantom barrier at $95 \%$ CL with $D_1$. Again, for the same model with $D_2$, the mean value of $w_{\rm de0}$ lies in the phantom region $(w_{\rm de0}< -1)$ at  $95 \%$ CL. Also, we notice a strong positive correlation between $w_{\rm de0}$ and the anisotropy parameter ($\Omega_{\rm \sigma0}$) in both cases, see Figure \ref{fig_desigma}. The upper bounds on $\Omega_{\rm \sigma0}$ are of the order $10^{-13}$ from $D_1$ and $D_2$. The constant EoS of the DE significantly affects the present matter density parameter and provides lower values for both models (as compared to models investigated in \cite{Yadav:2023yyb} with a cosmological constant) from both data. \\ 
We have obtained  higher values of the Hubble constant with $H_{\rm 0}=71.51\pm 0.83$ and $H_{\rm 0}=72.24\pm 0.82$ $\rm km\, s^{-1}\, Mpc^{-1}$ at 68\% CL for the $w$CDM and $w$CDM+$\Omega_{\rm \sigma0}$ models, respectively with $D_2$. The obtained value of $H_0$ in both the considered models from $D_2$ are consistent with SH0ES measurements ($H_0^{\rm R22}$), and thus the $H_0$ tension is relieved. Further, we notice that the inclusion of the anisotropy parameter in the $w$CDM model reduces the $H_0$ tension by $\sim 2\sigma$ and $\sim 1\sigma$ with $D_1$ and $D_2$, respectively.
Overall, we have shown how the anisotropic extensions of the standard cosmological model can influence the estimation of $H_0$, and contribute significantly to the resolution of the $H_0$ tension. 

\section*{Data Availability Statement} No data is associated with this manuscript.

\begin{acknowledgements}
The authors gratefully thank the reviewers for constructive comments and fruitful suggestions to enhance the quality of the manuscript. 
\end{acknowledgements}
\bibliographystyle{ieeetr}
\bibliography{refs}
\end{document}